\documentclass[%
reprint,
 amsmath,amssymb,
 aps, prl
]{revtex4-2}

\usepackage{xcolor}

\usepackage{svg}

\usepackage{amsmath}

\usepackage{blindtext}
\usepackage{float}
\usepackage{subfigure}

\usepackage{graphicx}
\usepackage{dcolumn}
\usepackage{bm}

\begin{document}

\preprint{APS/123-QED}

\title{Topological signatures of a p-wave superconducting wire through light}

\author{Frederick del Pozo and Karyn Le Hur}
\affiliation{CPHT, CNRS, Institut Polytechnique de Paris, Route de Saclay, 91120 Palaiseau, France}%
\date{\today}

\begin{abstract}
We show how the $\mathbb{Z}_{2}$ topological index of a one-dimensional topological p-wave superconductor can be revealed when driving with a classical vector potential i.e. an electromagnetic wave, through the
light-induced transition probabilities and the profile of the induced quasiparticles population. As a function of driving frequency $\omega$, it is possible to obtain a measure of this topological invariant from the resonance envelope classifying the two distinct topological phases of the short-range Kitaev wire. We propose to probe the topological phase transition in the model through the responses of the global capacitance in the presence of the light field and through the Josephson current between the wire and the proximity coupled bulk superconductor. The system may also be implemented on the Bloch sphere allowing alternative ways to measure the $\mathbb{Z}$ and $\mathbb{Z}_2$ topological invariants through circuit or cavity quantum electrodynamics. 
\end{abstract}

\maketitle
\emph{{\color{blue} Introduction .---}}
The realization and characterization of topological superconductors are attracting a lot of attention since the last decades. This is primarily due to their robustness against perturbations, and secondly due to zero-energy excitations confined to the edges of the system, guaranteed from the bulk-edge correspondence relating a mathematical \emph{invariant} to a physically observable edge excitation. In one-dimensional (1D) materials these zero-energy modes are localized at the edges, which can e.g. turn into Majorana fermions as a result of a magnetic impurity \cite{KLH2000}. 
The Kitaev wire \cite{KitaevO} is a famous prototype of a $p$-wave superconductor hosting a free Majorana fermion at each edge (MBS), leading to promising applications for noise-resilient quantum computing and as building blocks of superconducting circuits and hetero-structures \cite{MajRev1, Yang_2020}. The Kitaev wire is also related to the XXZ Ising chain via the Jordan-Wigner transformation, and as such has been realised experimentally also with evidence of the topological properties and phase transition \cite{QuantumIsingExp2022}. Majorana fermions can also occur in two-dots systems \cite{Delft,Flensberg} or in two Bloch-spheres' model \cite{KLHMajorana,MajoranaSpheres,HH}, such that the Kitaev wire remains at the heart of modern research. With the prominence of quantum technologies, the Kitaev wire with its various realisations remains very relevant as a platform to integrate both experimental and theoretical studies in quantum (information) physics.

Recent theoretical studies on the topological phase diagram of interacting Kitaev wires \cite{Schuricht} and hetero-structures thereof \cite{del2023, Herviou_2016, Yang_2020} have revealed a rich phase-diagram in the presence of strong couplings and interactions. 
Despite their robust nature, direct evidence of topologically non-trivial signatures is yet very challenging to obtain, motivating important current efforts. The Josephson effect \cite{Josephson62}, or more broadly the super-current between a superconductor and a metal \cite{Cohen62}, has been found to reveal signatures of topology in quantum wires. For example in SNS-type or Horse-shoe junctions, the current between the edges of topological superconductors obtains an additional $4\pi$-periodicity \cite{KitaevO,FuKaneCurrent} in the superconducting phase $\phi$, as opposed to the usual $2\pi$ one. Other proposals have addressed the critical current between the edges as a probe \cite{MappingmultibandMajorana14, Cayao17, Liu21}. The dynamical current susceptibility \cite{DmytrukTrif2018} in the presence of a time-dependent flux was also proposed as a protocol to probe MBS's. However, Andreev-bound-states (ABS) which arise inside the junction can both mimic zero-energy states \cite{DmytrukLoss2018} as well as topological signatures in the Josephson current \cite{Liu}. Despite numerous important efforts, measuring the topological invariant(s) in a 1D p-wave superconductor remains a challenge. 

The main goal of our letter is therefore to propose a simple way to reveal the topological invariant of the Kitaev wire and quantum phase transition through classical light, in the general sense of an electromagnetic wave. Through a junction between the wire and a proximity coupled substrate, we reveal characteristics of the $\mathbb{Z}_2$ topological index \cite{KitaevO,SatoAndo,hur2023topological} from geometry and the responses functions
associated to the light-induced inter-band transitions. From the map onto a quantum Bloch sphere describing the model in reciprocal (or momentum) space,  a link to the $\mathbb{Z}$ topological invariant is also possible similar to the Haldane model in two dimensions \cite{Roushan,hur2023topological,HH,del2023}. Finally we also discuss probing the topological phase transition e.g. from the Josephson effect and from the linear response of the global capacitance of the wire in the presence of an electromagnetic wave. In the low-frequency limit a clear signal is found, directly related to the bulk-gap closing at criticality. Through light, we then refer to electromagnetic signals in a general sense including radio and microwaves waves in the low-frequency domain. 

\emph{{\color{blue} The model .---}} We investigate the model of a 1D p-wave superconducting wire of spinless fermions, also known as the Kitaev wire. The Hamiltonian \cite{KitaevO} reads
\begin{equation}\label{Ham_c}
      H_{K} = \sum_{i} \left(-{\frac{\mu}{2}} c^{\dagger}_{i}c_{i} - t c^{\dagger}_{i}c_{i+1} +  \triangle e^{i\phi}c^{\dagger}_{i}c^{\dagger}_{i+1} + \text{h.c.} \right)
\end{equation}
where $t$ and $\Delta$ are the hopping and superconducting-pairing amplitudes with SC phase $\phi$ respectively, whilst $\mu$ is an on-site chemical potential. The pairing term is implicitly induced through proximity effect such that $\triangle=K\left\langle b\right\rangle$ with $b$ corresponding to Cooper pairs in a BCS substrate, and for simplicity we assume spin-polarized electrons to realize a p-wave symmetry. The model is general in the sense that it can be equivalently achieved through a Rashba spin-orbit coupling in nanowires with spin-1/2 electrons \cite{wires} and quantum spin chains \cite{QuantumIsingExp2022}. The references are ubiquitous, with e.g. reviews in
\cite{MajRev1,SatoAndo}. The Hamiltonian in \eqref{Ham_c} admits two topologically distinct phases, with quantum critical points (QCPs) at $\mu/t = \pm 2$ \cite{KitaevO, MajRev1}. The QCP belongs to the Ising universality class and is described by a free, chiral Majorana field \cite{Herviou_2016,del2023,KLH1999}. The topological phase of the Kitaev wire in Eq. \eqref{Ham_c} can be distinguished by a $\mathbb{Z}_2$ index \cite{KitaevO,SatoAndo}, similar to the Fu, Kane and Mele invariant \cite{KaneMele,FuKane}. Equivalently \cite{del2023}, they can also be characterized by a ``Chern number" \cite{Hur2022, HH, hur2023topological}. In the Bogoliubov-De Gennes (BdG-) representation $\psi^{\dagger}_{k} = \left( c^{\dagger}_{k}, c_{-k}\right)$, we find for \eqref{Ham_c}
\begin{equation}\label{Ham_Nambu}
    H_{K} = \frac{1}{2}\sum_{k\geq 0} \psi^{\dagger}_{k}  \mathcal{H}_{k} \psi_k \equiv \sum_{k\geq 0}\psi^{\dagger}_{k}
    \begin{pmatrix}\epsilon_{k} & \Delta_{k} \\ 
  \Delta^{*}_{k} & -\epsilon_{k}
\end{pmatrix}\psi_{k} ,
\end{equation}
with $\epsilon_{k} = -\left(\mu + 2t\cos\left(k\right)\right)$, $\Delta_{k} = 2i\triangle e^{i\phi} \sin\left(k\right)$ \cite{del2023} and Fourier transform $c_{j} = \frac{1}{\sqrt{N}}\sum_{k \in BZ} e^{ikj}c_{k}$. The length is $L = Na$, with lattice spacing $a=1$ and $N$ the number of sites.
The Hamiltonian \eqref{Ham_Nambu} is diagonal in the particle-hole symmetric (PHS) quasi-particle (QP) basis defined by $\eta^{+}_{k} = u_{k} c_{k} + v_{k}c^{\dagger}_{-k}$ and $\eta^{-}_{k} = v^{*}_{k} c_{k} - { u^{*}_{k}}c^{\dagger}_{-k}$, with $u_{-k} = -u_{k}$ and also $|u_{k}|^{2} + |v_{k}|^{2} = 1$. We introduce a representation on the Bloch-sphere through \cite{hur2023topological, Herviou_2016, del2023}
\begin{equation}\label{BS_mapping}
\begin{aligned}
    \cos \left(\theta_k\right)  =&\frac{2t \cos \left(k\right)+\mu}{E\left(k\right)} =-\frac{\epsilon_k}{E\left(k\right)}\\ 
    \sin \left(\theta_k\right) e^{-i \varphi_{k}} =&-\frac{2i \Delta e^{i \phi}\sin \left(k\right)}{E\left(k\right)} = -\frac{\Delta_{k}}{E\left(k\right)},
\end{aligned}
\end{equation}
with spectrum ${E(k) = \sqrt{\epsilon^{2}_{k} + |\Delta_{k}|^2}}$ and $\varphi_{k}=-\phi+\frac{\pi}{2}$. Additionally $u_{k} = -ie^{-i\phi}\sin\left(\frac{\theta_{k}}{2}\right)$ and $v_{k} = \cos\left(\frac{\theta_{k}}{2}\right)$. We equivalently have $H_{K} = \sum_{k\geq 0} E\left(k\right)\left(\eta^{+\dagger}_{k}\eta^{+}_{k} - \eta^{-\dagger}_{k}\eta^{-}_{k}\right)$, such that for $E(k) > 0$ the GS is determined by the vacuum of $\eta_k^{+}$ particles or equivalently (by PHS) a fully occupied $\eta^{-}_{k}$ band, defined as \cite{del2023}
\begin{equation}\label{BCS_WF}
\begin{aligned}
    &|\mathrm{BCS} \rangle = \left(\delta_{\mu<-2 t}+\delta_{\mu>-2 t} c_0^{\dagger}\right)
\Big(\delta_{\mu<2 t}+\delta_{\mu>2 t} c_\pi^{\dagger}\Big) \\
&\times 
\prod_{k>0}^{k<\frac{\pi}{a}}\left(\sin \left(\frac{\theta_k}{2}\right)-i e^{i \phi} \cos \left(\frac{\theta_k}{2}\right)c^{\dagger}_{k}c^{\dagger}_{-k}\right)|0\rangle.
\end{aligned}
\end{equation}
The $\eta^{+}$'s annihilate the BCS state and are given by
\begin{equation}\label{qp_plus}
\begin{aligned}
    \eta^{+}_{k} = &-ie^{-i\phi}\sin\left(\frac{\theta_{k}}{2}\right)c_{k} +\cos\left(\frac{\theta_{k}}{2}\right)c^{\dagger}_{-k}.
    \end{aligned}
\end{equation}
From now on we write $\eta^{+}_{k} = \eta_{k}$.

\emph{\color{blue} Josephson Current.---} Coupling two superconductors results in the flow of a current through a junction \cite{Cohen62, Josephson62, Wallace64}, known as the Josephson effect. We show how the phase transition is visible from the {\it bulk} current $\mathcal{J}$ between the wire and a proximized BCS substrate. {\color{black} It is defined via the continuity equation $e\frac{\partial \mathcal{N}}{\partial t} - \mathcal{J} = 0$, with particle number ${\cal N}$ \cite{Cohen62, Josephson62}.} Fixing $\hbar=\frac{h}{2\pi}=1 = e$, we find
\begin{equation}
\mathcal{J} = \frac{\partial {\cal N}}{\partial t} = i[H_K,{\cal N}] \equiv \mathcal{J}_{\Delta}.
\end{equation}
{ The current within the wire resulting from the hopping of electrons is $\frak{J}_t  = -i t \sum_i \left(c^{\dagger}_{i+1}c_i - c^{\dagger}_i c_{i+1}\right)$, and vanishes on the BCS ground state.}
The superconducting term $\mathcal{J}_{\Delta}$ on the other hand drives a super-flow between the wire and the substrate, defined for periodic boundaries, as
{\color{black}
\begin{equation}
    \mathcal{J}_{\Delta} = \mathcal{J}_{\Delta}^{+}\left(2e\right) + \mathcal{J}^{-}_{\Delta}\left(2e\right),
\end{equation}
with $\mathcal{J}^{+}_{\Delta} = 2i\Delta e^{i\phi} \sum_i c^{\dagger}_{i+1} c^{\dagger}_i$ and $\mathcal{J}^{-}_{\Delta} = \left(\mathcal{J}^{+}_{\Delta}\right)^{\dagger}$.
} On the BCS ground state with $\left\langle c^{\dagger}_k c^{\dagger}_{-k}\right\rangle = \frac{i}{2}\sin\theta_k$ \cite{Herviou_2016, del2023}, the current for a wire of length $L$ is 
\begin{equation}\label{JosephCurrent}
\langle \mathcal{J}\rangle = \langle\mathcal{J}_{\Delta}\rangle =  \frac{2L\Delta}{\pi}\sin\left(\phi\right) \int_0^\pi dk \sin \left(k\right) \sin(\theta_k),
\end{equation}
where we took the continuum limit $\sum_{k = 0}^{\pi} \longrightarrow \frac{1}{\alpha}\int_{0}^{\pi}\text{d}k$, with $\alpha = \frac{2\pi}{L}$. The notable difference between the current \eqref{JosephCurrent} between two p-wave SCs, and the current between two $s-$wave SCs, is the $\sin\left(k\right)$ in the integrand \cite{Wallace64}. 

\begin{figure}[h!]
\includegraphics[width=\linewidth]{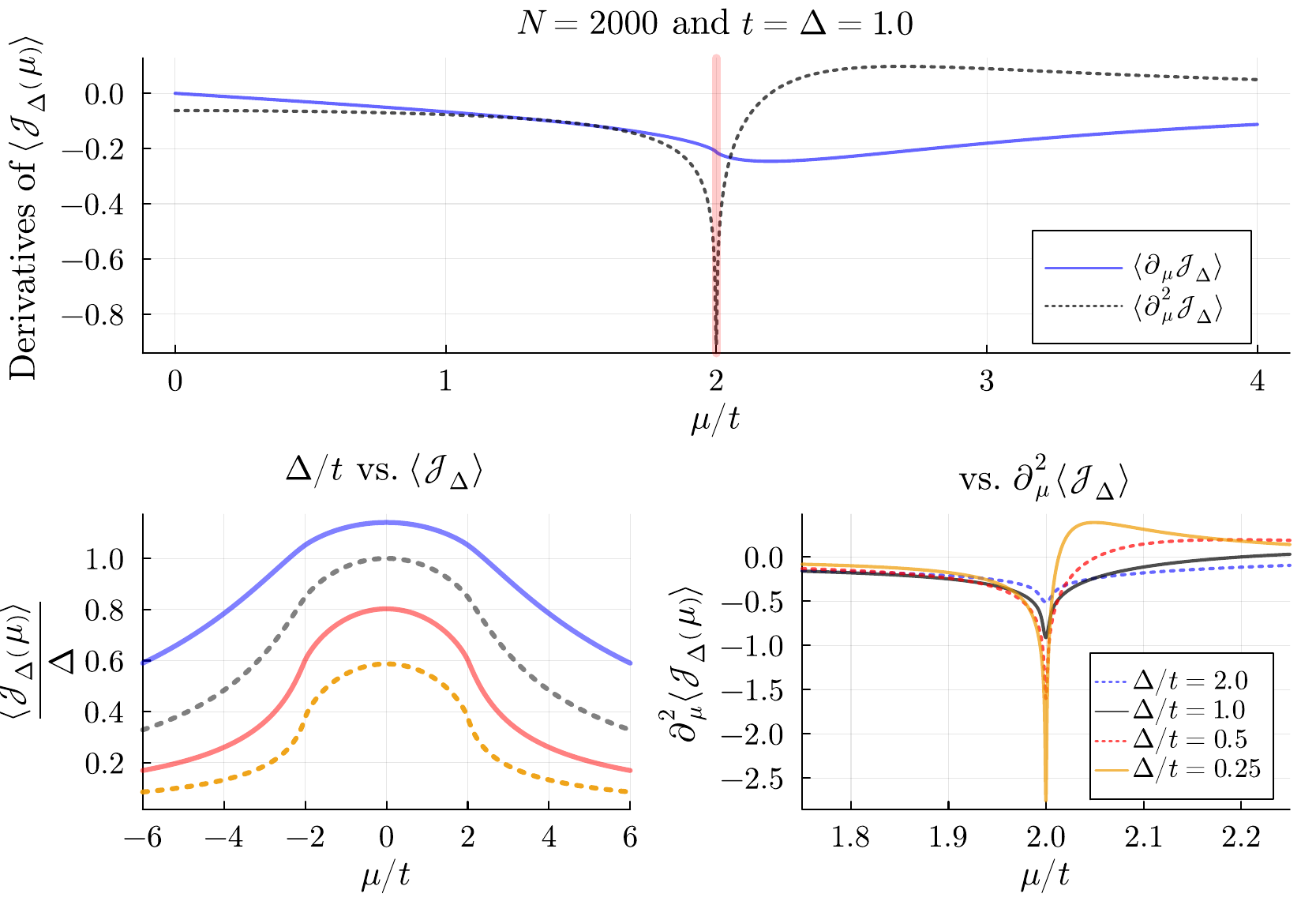}
\caption{Josephson currents and derivatives at $\phi = \pi/2$ and $L = 2000a$ with $a = 1.0$. (Upper) { QCPs} are revealed through the diverging slope of $\partial_{\mu}\langle  \mathcal{J}_{\Delta}\rangle$. (Lower) Currents and second derivatives at various $\Delta/t$. The transition becomes visibly pronounced in the low $\Delta/t$ limit. The results are shown in units of $[L]$.}
\label{fig:Josephson}
\end{figure}

In Fig. \ref{fig:Josephson} above, we show the Josephson current accross the phase transition for various values of $\Delta/t$. Notably, we find a divergence of the second derivative of $\mathcal{J}$ at the QCP. This is a direct consequence of the gap-closing. As is presented in more detail in the supplementary material (SM) \cite{SM}, we find from Eq. (\ref{JosephCurrent}) the second derivative of the Josephson current
\onecolumngrid
\begin{equation}
\frac{\partial^2 \langle \mathcal{J}\rangle}{\partial \mu^2} = -\frac{L\Delta}{2\pi} \sin \phi \int_0^{\pi} \frac{\sin^2\left( k\right)}{E(k)^3} dk 
{+}3\frac{L\Delta}{\pi} \sin \phi \int_0^{\pi} \sin^2 \left(k\right) \frac{\epsilon_k^2}{E(k)^5} dk.
\end{equation}
\twocolumngrid
As we show in the SM \cite{SM}, the first integral diverges close to $0$ or $\pi$ when $\mu/t = \pm 2$. 
Through this divergence of  $\partial^{2}_{\mu}\langle{\cal J}\rangle$  at  $\mu/t = \pm 2$, the current provides interesting insights into the bulk properties of Kitaev wires compared to probes such as bipartite fluctuations, quantum Fisher Information density \cite{Fisherinformation} or the dynamical susceptibility in SQUID junctions \cite{DmytrukTrif2018}.

\emph{\color{blue} Response to classical light.--- }{\color{black} Various protocols have suggested measuring the topological properties of a wire }through a cavity field in different geometries \cite{Dmytruk16,Takis}.  The same charge fluctuations and dynamical susceptibility addressed the response of the 
Su-Schrieffer-Heeger (SSH) model \cite{Tal,Boulder} and the response of a 1D p-wave superconducting wire \cite{Dmytruk15} in circuit quantum electrodynamics. For the SSH chain {\color{black} which presents a similar 
$\mathbb{Z}$ invariant,} this response also allowed to reveal the Zak phase (or winding number) \cite{Tal,Boulder}.
{\color{black} Whereas various protocols were suggested and then measured for the SSH model  \cite{Tal,Boulder,StJean},
to the best of our knowledge measuring the topological invariant for the p-wave superconducting wire remains a challenge.} Therefore we propose to measure the $\mathbb{Z}_2$ topological number through the response to classical light. {\color{black} Aligning the direction of the classical vector potential $\vec{A}$ to the direction associated to $\mathcal{J}_{\Delta}$ yields
\begin{equation}
    \delta H\left(t\right) = -\mathcal{A}\left(t\right) \mathcal{J}_{\Delta}^{+}  - \mathcal{A}^{*}\left(t\right)\mathcal{J}_{\Delta}^{-},
\end{equation}
in direct analogy to the 2D light-matter interaction with circularly polarized light \cite{KleinGrushinLeHur_Stoch, Hur2022} and from mesoscopic quantum electro dynamics \cite{Cottet_2017, CottetKontosDou_SC_Peierls},
where through Peierls transformation, the light-matter interaction dresses the SC phase $\phi$.}  By introducing the quasi-particle operators $O_{k}^{z} = \left(\eta^{\dagger}_{k}\eta_{k} - \eta_{-k}\eta^{\dagger}_{-k}\right)$, $O^{+}_{k} = \eta^{\dagger}_{k}\eta^{\dagger}_{-k}$ and $O^{-}_{k} = \left(O^{+}_{k}\right)^{\dagger}$,
\onecolumngrid
{
\begin{eqnarray}\label{Oops}
    \delta H\left(t\right) = -4\Delta\sum_{k = 0}^{\pi}\sin(k) \left[i\alpha_{k} O^{z}_{k}\left( \mathcal{A}(t) - \mathcal{A}^{*}(t)\right) +\left(\beta_{k} \mathcal{A}(t) + \tilde{\beta}^*_{k}\mathcal{A}^{*}(t)\right) O^{+}_{k} + \left(\tilde{\beta}_{k}\mathcal{A}(t) + \beta^*_{k}\mathcal{A}^{*}(t)\right)O^{-}_{k} \right],
    \end{eqnarray}
    }
 \twocolumngrid
\hskip -0.3cm
with definitions $2\alpha_{k} = -\sin\left(\theta_{k}\right)$, $\beta_{k} = e^{-i\phi}\sin^{2}\left(\frac{\theta_{k}}{2}\right)$ and $\tilde{\beta}_{k} = e^{i\phi}\cos^{2}\left(\frac{\theta_{k}}{2}\right)$.
Therefore, classical light both dresses the BCS vacuum with $O^{z}_{k}|BCS\rangle = |BCS\rangle$, and raises quasi-particles and quasi-holes. {\color{black} By coupling to linearly polarized light we now show how to measure the $\mathbb{Z}_{2}$ index of the Kitaev wire via QP transition rates.}

\emph{\color{blue} Topological Signatures.---}{\color{black} We now consider the response to a coherent signal with tunable frequency $\omega$. The vector potential describing such an electro-magnetic wave is given by $\mathcal{A}\left(t\right) = A_{0}e^{i\omega t}$. To leading order, the perturbation can be treated in the interaction-picture $ |BCS\left(t\right)\rangle \approx \left[\mathbb{I} -\frac{i}{\hbar}\int_{0}^{t} \text{d}\tau \delta H_{I}\left(\tau\right) \right]|BCS\rangle$
with subscript $I$ denoting the interaction-picture representation $O_{I}\left(t\right) = e^{iH_{0}t}O\left(t\right)e^{-iH_{0}}$.} We obtain
\onecolumngrid
\begin{equation}\label{time_evol_BCS_rate}
|B C S(t)\rangle=\mathfrak{a}\left(t\right)|B C S\rangle-i\sum_{k = 0}^{\pi} B_{0} \sin \left(k\right) \eta_k^{\dagger} \eta_{-k}^{\dagger} \int_0^t \left [ e^{-i\phi} \sin ^2 \left(\frac{\theta_{k}}{2}\right)  e^{i \omega^{+}_{k} \tau} +  e^{i\phi}\cos^2 \left(\frac{\theta_{k}}{2}\right)  e^{-i \omega^{-}_{k} \tau}  \right] d \tau|B C S\rangle.
\end{equation}
\twocolumngrid
We used the following definitions $\omega^{\pm}_{k} = \omega \pm 2E\left(k\right)$, $B_{0} = 4\frac{\Delta A_{0}}{\hbar} \ll \hbar^{-1}$ to simplify, and also write
\begin{equation}
\mathfrak{a}\left(t\right) = \left(1 + 2B_{0}\int_{0}^{t}\text{d}\tau\sin\left(\omega \tau\right) \sum_{k = 0}^{\pi} \sin\left(k\right) \alpha_{k}\right).
\end{equation}
{\color{black} The topology of the Kitaev wire is characterized by a \emph{winding number} $m$. One way to define it is by the amount of times the Bloch sphere map covers the entire sphere $\mathbb{S}^{2}$. For $m = 0$ we have $\theta_{k} < \frac{\pi}{2}$ and within the topological phase, for $m =1$ the angle $\theta_{k} \in \left[0,\pi\right]$. Therefore, all the information on topology is encoded in the trigonometric function $\sin^{2}\left(\frac{\theta_{k}}{2}\right) - \cos^{2}\left(\frac{\theta_{k}}{2}\right)$, which is zero at $\theta_{k} = \frac{\pi}{2}$.}

We propose to measure the $\mathbb{Z}_{2}$ index through the light-induced transition of an $\eta$ quasi-particle into the upper-band, ie. $\sim \eta^{\dagger,+}_{k}\eta^{-}_{k}$. With PHS this defines the excited state $|ES\rangle \equiv \sum_{q = 0}^{\pi} \eta^{\dagger}_{q}\eta^{\dagger}_{-q}|BCS\rangle$, and transition probability $\mathcal{P}\left(t\right) = \left|\sum_{q= 0}^{\pi} \left\langle BCS| \eta_{-q}\eta_{q} |BCS\left(t\right)\right\rangle\right|^{2}$. {\color{black}This measurement is ``blind" to weak disorder $\delta\mu\left(x\right)$, which either acts on the poles $k = 0$ and $k = \pi$ or results in excitations $\sim \eta^{\dagger}_{k + q}\eta_{k}$ \cite{SM}.} By introducing a small damping term 
$\epsilon > 0$ into the vector potential, for $t \gg 1/\epsilon$ we find $\int_{0}^{t \gg 1/\epsilon}\text{d}\tau e^{i\left(\pm \omega^{\pm}_{k} + i\epsilon\right)\tau} \approx \frac{1}{\pm i\omega^{\pm}_{k} -\epsilon}$
which becomes resonant at $\omega^{\pm}_{k} =0 $ and has a finite width/height proportional to $\epsilon$. Physically, this can be related to a life-time. At $t = \infty$ we obtain $\mathcal{P}\left(\infty\right)$ as
\begin{equation}\label{envelope}
\begin{aligned}
   \frac{\mathcal{P}\left(\infty\right)}{B_{0}^{2}} =  \left|\sum_{k} \sin\left(k\right) \left[\frac{e^{-i\phi}\sin^{2}\left(\frac{\theta_{k}}{2}\right)}{ \omega^{+}_{k} + i\epsilon} - \frac{e^{i\phi}\cos^{2}\left(\frac{\theta_{k}}{2}\right)}{\omega^{-}_{k} - i\epsilon}  \right] \right|^{2}.
\end{aligned}
\end{equation}
We measure the response at resonance at $\omega^{\pm}_k=0$ with two distinct envelopes depending on the sign of $\omega$. 

\begin{figure}[ht]
\vskip 0.3cm
\includegraphics[width=\linewidth]{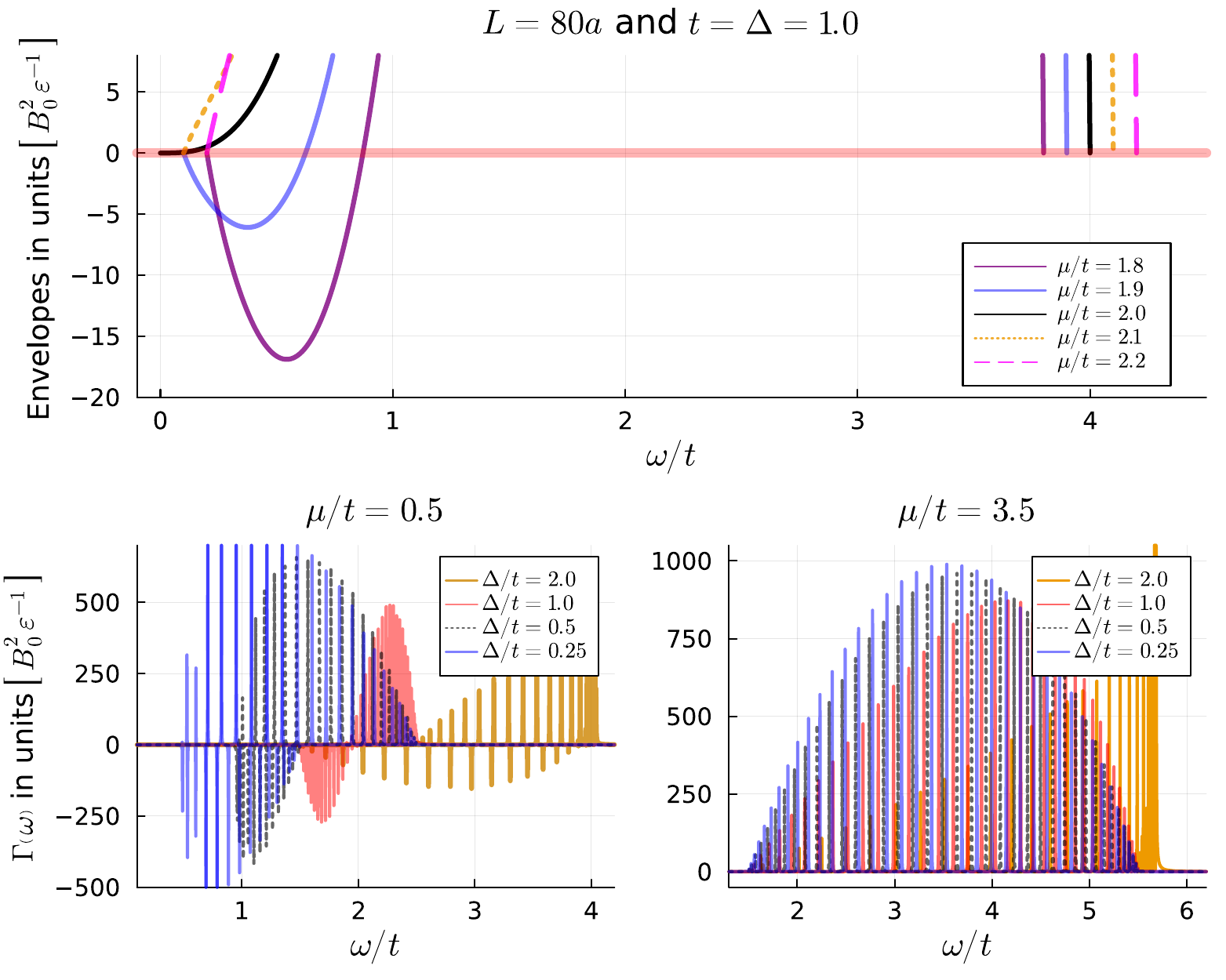}
\caption{(Upper) Resonance envelopes of \eqref{envelope} at $t = \Delta = 1.0$. The QPT is then distinguished by the additional zero in the central domain. (Lower) Resonance peaks at $\mu/t = 0.5$ in the topological and $\mu/t = 3.5$ in the trivial phase at different values of $\Delta/t$. The overall shape of the envelope is considerably different to $\Delta/t = 1.0$, however $\Gamma\left(\omega\right) \geq 0$ is still guaranteed for trivial winding numbers $\nu = 0$, whilst $\Gamma\left(\omega\right) < 0$ is possible in the topological regime. $\epsilon = 10^{-3}$ and $\Delta \omega/t = 5.0\cdot 10^{-4}$ for the resonance peaks. Envelopes scaled by global constant (here $b = 6.3$) to match peak heights qualitatively. The system size is $L = 80a$ with $a = 1.0$, to guarantee visually suitable resolution of the resonance peaks.}
\label{fig:res_env}
\end{figure}

The phase $\phi$ is set to zero for simplicity, and as a result of the absolute values any dressing of the SC phase due
to additional Peierls transformation will at most lead to oscillations of the envelopes.
We define $\mathcal{P}_{+}\left(\infty\right) = \mathcal{P}\left(\infty; \omega >0 \right)$, and $\mathcal{P}_{-}\left(\infty\right) = \mathcal{P}\left(\infty; \omega <0 \right)$, and propose to introduce
\begin{equation}\label{Gamma}
\Gamma\left(\omega\right) =  \mathcal{P}_{+}\left(\infty\right) - \mathcal{P}_{-}\left(\infty\right).
\end{equation}
For $t = \Delta$ this reduces to exactly to a $\mathbb{Z}_{2}$ invariant, as here $E\left(k\right)$ is injective and the envelope of $\Gamma\left(\omega\right)$ presents exactly three zeros for $|\mu/t |<2$: when $\sin(k)=0$ i.e. at $k=0,\pi$ and additionally when $\sin \frac{\theta_k}{2}=\cos\frac{\theta_k}{2}$ corresponding to $\theta_k=\frac{\pi}{2}$ with $k$ solution of $\omega^{\pm}_k=0$; see Fig. \ref{fig:res_env}. The trivial phase for $\mu>2t$ or $\mu<-2t$ is characterized through the two zeros of the same function at $k=0$ and $\pi$. Therefore, we propose to measure the invariant $(-1)^{\zeta}$, where $\zeta$ measures the number of zeros of $\Gamma(\omega)$.

This makes a link with the ``Chern marker" introduced in \cite{Hur2022,hur2023topological} measurable from the pseudo-spin $S_z=(c^{\dagger}_k c_k - c_{-k}c^{\dagger}_{-k})$. The $\mathbb{Z}$ topological number can then be formulated as $C=\frac{1}{2}(\langle S_z(0)\rangle - \langle S_z(\pi)\rangle)$ on the Bloch sphere with $S_z=S_z(\theta_k)$ \cite{del2023}. At the particular angle $\theta_k=\frac{\pi}{2}$ we can write the identity $2\cos^4 \frac{\theta_k}{2} = 2\sin^4 \frac{\theta_k}{2} = C^2 - \frac{1}{2}$ within the topological phase for $|\mu|<2t$. 
The fact that $\theta_k\in [0;\frac{\pi}{2}]$ for $\mu=\pm 2t$ also reveals that the topological invariant jumps from $1$ to $0$ corresponding then to $C=\frac{1}{2}$ or to a winding number on half a sphere \cite{hur2023topological, del2023}, i.e. $\langle S_z\left(\frac{\pi}{2}\right)\rangle=0$. 

{\color{black} $E\left(k\right)$ is no longer injective when $t \neq \Delta >0$. Therefore, $\omega^{\pm}_{k} = 0$ for two wavelengths in some cases, which \emph{in the topological region only} may lead to negative resonance peaks for small $\epsilon$ and lower $k-$space resolution, \emph{cf.} bottom-left panel of Fig. \ref{fig:res_env}. However, we verified numerically that these wash out for larger $\epsilon$ or finer grids, and the resulting resonances again develop exactly one additional zero in the topological phase. }

At this stage, it is interesting to observe the relation with the observable $\langle BCS(t)| \sum_k\eta^{\dagger}_k \eta_k |BCS(t)\rangle=\sum_k \langle BCS(t) | \eta^{\dagger}_k \eta_k |BCS(t)\rangle = \sum_k {\cal P}_k(t)$ where
\begin{equation}
\frac{{\cal P}_k(\infty)}{B_0^2} =  \left|\sin\left(k\right) \left[\frac{e^{-i\phi}\sin^{2}\left(\frac{\theta_{k}}{2}\right)}{ \omega^{+}_{k} + i\epsilon} - \frac{e^{i\phi}\cos^{2}\left(\frac{\theta_{k}}{2}\right)}{\omega^{-}_{k} - i\epsilon}  \right] \right|^{2}.
\end{equation}
Measuring the number of QP in the upper band then encodes similar information as in Eq. (\ref{envelope}) when light is at resonance, \emph{cd.} the number of zeros in Fig. (\ref{fig:res_env}) (top). Last but not least, when $\omega\rightarrow 0$, Eq. (\ref{envelope}) also probes the BCS pairing strength \cite{SM}. Driving QP transitions with classical light therefore enables numerous potential measurement protocols.

\emph{\color{blue} Capacitance probe .---}
Lastly we demonstrate that the topological phase transition can be revealed from the total charge $\mathcal{Q}\left(t\right)$ at time $t$ in the presence of the light field $\mathcal{A}(t)$. This could be measured from the capacitance $C = \mathcal{Q}/\mathcal{V}$, where $\mathcal{V}$ is the voltage difference between the wire and the probe. The response to linear order is given by the Kubo formula \cite{Kubo}, and we measure $\Delta \mathcal{Q}_{L}\left(t\right) =  -i\int_{0}^{t}\text{d}\tau \mathcal{A}\left(\tau\right)\left\langle\left[\mathcal{Q}_{I}\left(t\right), \mathcal{J}_{I}\left(\tau\right)\right]\right\rangle_{0}$ in the interaction representation with $|\psi\left(0\right)\rangle = |BCS\rangle$. For the sake of clarity, we present the detailed calculation in the Supplementary Material \cite{SM}. 
{

For $\phi = 0$, the current operator $\mathcal{J}_{\Delta}$ is given in terms of the quasi-particle operators $O^{\pm}$ and $O^{z}$ as $
\mathcal{J}_{\Delta} = {\color{black}4}\Delta \sum_{q} \sin\left({q}\right)\left(O_q^{+} + O_q^{-}\right)$
whilst the charge reads
\begin{equation}
\mathcal{Q} = - \sum_{k}\left( \cos\left(\theta_{k}\right)O^{z}_{k} + i\sin\left(\theta_{k}\right)\left(O^{+}_{k} - O^{-}_{k}\right)\right).
\end{equation} \\

Due to $O_{q}^{-}|BCS\rangle = 0$, only the $\left\langle O_{q}^{-}O_{k}^{+} \right\rangle_{0}\sim \delta\left(k- q\right)$ contributes non-trivially to the response $\Delta \mathcal{Q}_{L}\left(t\right)$. Together with $\omega^{\pm}_{k} = \omega \pm 2E\left(k\right)$, we find the formula
\onecolumngrid
\begin{equation}\label{charge_time}
\begin{aligned}
       \Delta\mathcal{Q}_{L}\left(t\right)=&\ {\color{black} -} i\frac{B_{0}L^{2}}{4\pi^{2}}\int_{0}^{\pi}\text{d}k\int_{0}^{\pi}\text{d}q \int_{0}^{t}\text{d}\tau\ \sin\left(q\right)\sin\left(\theta_{k}\right)\cos\left(\omega \tau\right) \left\langle  i O^{-}_{q} e^{iH_{0}\left(t - \tau\right)} O^{+}_{k} - i O_{k}^{-} e^{-iH_{0}\left(t - \tau\right)} O^{+}_{q}\right\rangle_{0} \\ 
      =& {\color{black} +}\frac{B_{0}L}{2\pi}\int_{0}^{\pi}\text{d}k \sin\left(k\right)\sin\left(\theta_{k}\right) \left[ \frac{\sin\left(\omega t\right)  - \sin\left(2E(k)t\right)}{\omega^{+}_{k}} + \frac{\sin\left(\omega t\right)  + \sin\left(2E(k)t\right)}{\omega^{-}_{k}}\right].
\end{aligned}
\end{equation}
There are two competing oscillations in \eqref{charge_time} above, and we perform an additional time-average over the domain $t \in \left[0, \frac{2\pi}{\omega}\right]$ to remove the oscillations in $\omega t$.
}
Following the same approach as above by fixing $\phi=0$ for simplicity and for weak $B_0$
such that the topological phase is robust, we find the frequency dependent response
\onecolumngrid
\begin{equation}\label{tildeQ}
     \tilde{\mathcal{Q}}_{L}\left(\omega\right) = {\color{black}-}\frac{1}{T}\int_{0}^{T = \frac{2\pi}{\omega}}\Delta \mathcal{Q}_{L}\left(t\right)
       = - \frac{B_{0}L\omega}{4\pi^{2}}\int_{0}^{\pi}\text{d}k \frac{\sin\left(k\right)\sin\left(\theta_{k}\right)}{2E\left(k\right)}\sin^{2}\left(\frac{2\pi E\left(k\right)}{\omega}\right)\frac{4E\left(k\right)}{\omega^{2} - 4E^{2}\left(k\right)}.
\end{equation}

\twocolumngrid

\begin{figure}[ht]
\includegraphics[width=\linewidth]{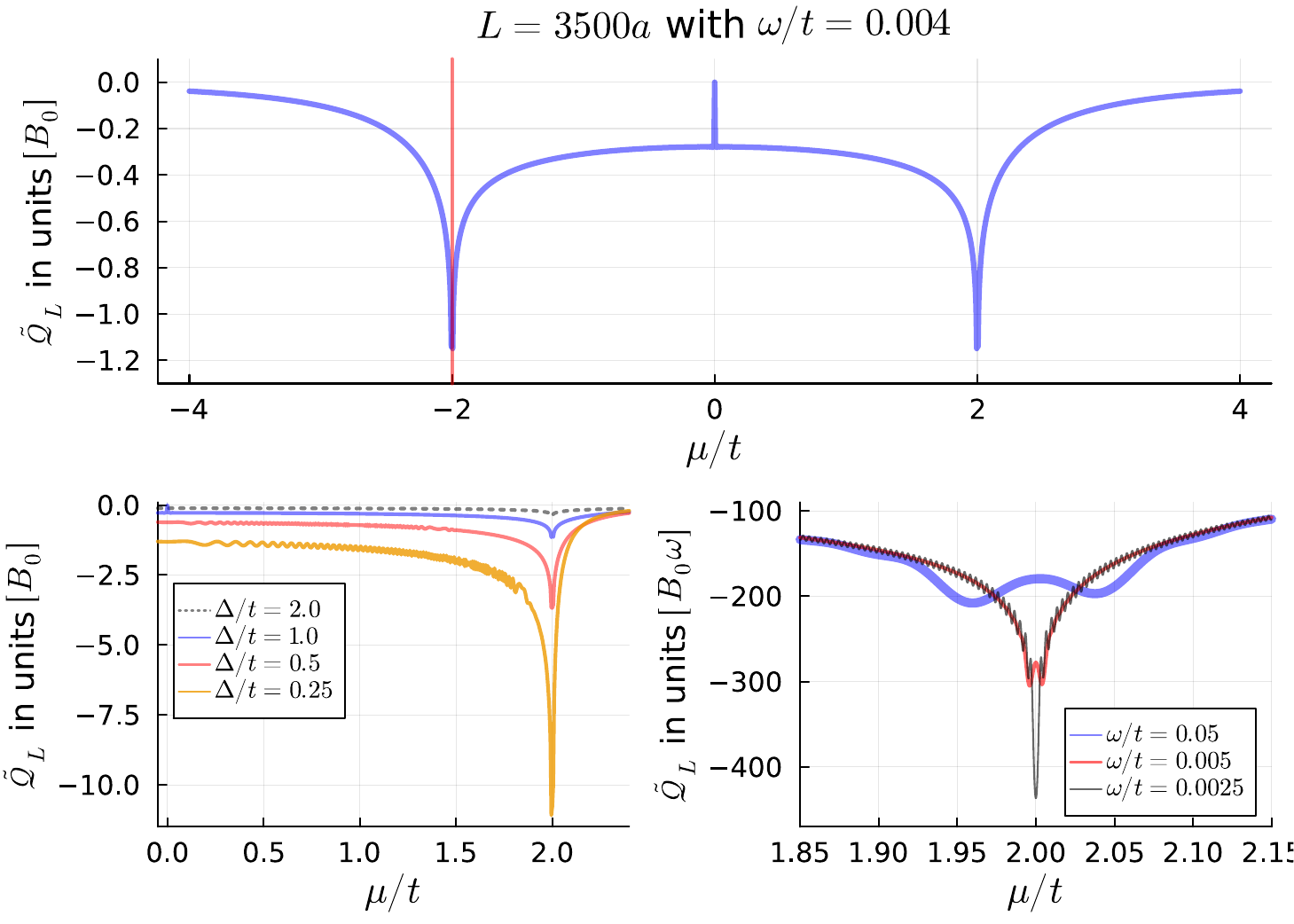}
\caption{Linear response of the total charge (number) operator $\mathcal{Q}$. (Lower right) Double peak around $\mu = 2t$ can be seen for increasing $\omega/t$, reflecting the resonances in \eqref{tildeQ} in the gapped phases. (Lower left) Dependence on $\Delta$, with a zero response at $\mu/t = 0.0$ only at $\Delta/t = 1.0$. We used $L = 3.5\cdot 10^{3}a$ and $\omega/t = 4.0 \cdot10^{-3}$, and $t = \Delta = 1.0$ unless otherwise stated.}
\label{fig:TildeQ}
\end{figure}

At low frequency $\omega/t \ll 1$, $\tilde{Q}_{L}\left(\omega\right)$ can additionally detect the QPT at $\mu = \pm 2t$, as seen in the upper panel of Fig. \ref{fig:TildeQ}. At comparatively larger values of $\omega$ a splitting of the peaks occurs, see lower-right panel. This comes from the $\frac{1}{\omega \pm 2E\left(k\right)}$ resonance terms. 
For $t = \Delta = 1.0$ we have a vanishing response at $\mu/t = 0$. In this limit $E\left(k\right) = t$ and $\theta_{k} = k$ \cite{del2023}, and we find analytically a scaling 
$\tilde{Q}_{L}\left(\mu = 0\right) \sim \text{sinc}^{2}\left(\frac{2\pi t}{\omega}\right)$. For $\Delta\rightarrow 0$, since $\theta_{k}\rightarrow 0$, we also verify that the result remains identical.
At $\mu = -2t$, we find with $\theta_{k} = k/2$ \cite{del2023} that there is a sharp dip in the response of $\tilde{Q}_{L}$ at low frequency (with a finite value). The dip becomes more pronounced for smaller $\Delta$ revealing that $E^{-2}(k)\sim (\Delta^2 k^2)^{-1}$ when $k\rightarrow 0$.

\emph{\color{blue} Conclusion .--- } We have presented several methods through light to reveal both the topological invariant(s) as well as the topological phase transition of a one-dimensional p-wave superconducting wire. We hope that our work can then give further insight on revealing the topological matter with electromagnetic waves. As a remark, we also find it useful to add that the presence of Majorana fermions (at zero energy) e.g. will not modify the finite-frequency light-induced transition probabilities and the capacitance measure (in the limit where $E(k)\rightarrow 0$). Also, at the topological transition, Majorana fermions are located at $k=0$ and $k=\pi$ which then correspond to two zeros of the function in Eq. \eqref{Gamma}.  

\emph{\color{blue} Acknowledgements .--- }
The authors would like to thank the Center for Theoretical Physics (CPhT) at Ecole Polytechnique and the CNRS for their continued support. Frederick del Pozo would in particular also like to thank the Ecole Doctorale of IP Paris, as well as the larger group and members of CPhT. Many thanks also go to Sariah al Saati, and espescially to Olesia Dmytruk for the many insightful and rewarding discussions throughout the project.  This  work  was  supported  by  the  french ANR BOCA and the Deutsche Forschungsgemeinschaft (DFG), German Research Foundation under Project No.277974659.


\bibliography{apssamp}

\newpage

\newpage

\onecolumngrid
{\color{blue}
\section*{Supplementary Material: Topological signatures of a p-wave superconducting wire through light}}

Here, we present additional information on the behavior of the Josephson current for the p-wave superconducting wire at the topological phase transition. We also show how Eq. (14) in the Letter probes the effective
BCS pairing strength. We also present the derivation of the capacitance measure at low frequency.\\

{\color{blue}\subsection{Josephson Current and Topological Quantum Phase Transition}
}
Here, we show how the Josephson current probes the topological phase transition of a one-dimensional p-wave topological superconducting Kitaev wire.  
Mathematically, from the definitions in Eq. (3) in the article we obtain
\begin{equation}
\frac{\partial \langle \mathcal{J} \rangle}{\partial \mu} = \frac{L\Delta}{\pi} \sin \phi \int_0^{\pi} \frac{\sin^2 k}{E(k)^3} \epsilon_k dk
\end{equation}
For $t=\Delta$, at $\mu=2t$ we obtain
\begin{equation}
\frac{\partial \langle \mathcal{J} \rangle}{\partial \mu} = - \frac{2}{3\pi} L\Delta \sin \phi
\end{equation}
and for $\mu=-2t$, 
\begin{equation}
\frac{\partial \langle \mathcal{J} \rangle}{\partial \mu} =  \frac{2}{3\pi} L\Delta \sin \phi
\end{equation}
which agrees with numerical results. For $\mu=0$, we can also verify that this integral is zero. In a similar way, we obtain
\begin{equation}
\frac{\partial^2 \langle \mathcal{J}\rangle}{\partial \mu^2} = -\frac{L\Delta}{2\pi} \sin \phi \int_0^{\pi} \frac{\sin^2\left( k\right)}{E(k)^3} dk 
{+}3\frac{L\Delta}{{2}\pi} \sin \phi \int_0^{\pi} \sin^2 \left(k\right) \frac{\epsilon_k^2}{E(k)^5} dk.
\end{equation}
The first integral diverges close to $0$ or $\pi$ when $\mu/t = \pm 2$. To see this, we first substitute $E^{2}\left(k\right) = \epsilon_k^2 +| \Delta_{k}|^2$, and make the observation that $E\left(k\right) = 0$ for $k = 0$ or $\pi$ only. This is also true for $\Delta \neq t$, such that we focus in the remainder on the special case $t= \Delta = 1.0$ and $\phi = \pi/2$. Then
\begin{equation}
 -\frac{L}{2\pi} \int_0^{\pi} \frac{\sin^2\left( k\right)}{E(k)^3} dk \approx  -\frac{L}{2\pi} \int_0^{\pi} \frac{\sin^2\left( k\right)}{\left(\left(\mu +2t\cos\left(k\right)\right)^{2} + 4\sin^{2}\left(k\right)\right)^{\frac{3}{2}}} dk
\end{equation}
Close to $\mu = \pm 2t$, ie. $|\delta\mu| = |\mu \pm 2t| \ll 1$, the gap closes either at $k = 0$ or $k = \pi$. Fixing $\delta\mu < 0$ we have the gap-closing around $k = 0$ and hence the divergence will be dominated by a ``disk" of radius $\epsilon$ around $k = 0$. Here We may approximate $\epsilon_{k} \approx \delta\mu - \frac{k^{2}}{2}$ as well as $\sin^{2}\left(k\right) \approx k^{2}$. Therefore, we find
\begin{equation}
\begin{aligned}
       -\frac{L}{2\pi} \int_0^{\pi} \frac{\sin^2\left( k\right)}{E(k)^3} dk    \approx & -\frac{L}{2\pi} \int_0^{\epsilon} \frac{k^{2}}{\left(\left(\delta\mu -\frac{k^{2}}{2}\right)^{2} + 4k^2\right)^{\frac{3}{2}}} dk + \text{finite terms} \\ 
       \approx & -\frac{L}{2\pi}\lim_{\alpha \longrightarrow 0 }\left[ \int_{\alpha}^{\epsilon} \frac{k^{2}}{\left( \frac{k^{4}}{4} - k^{2}\delta\mu + 4k^{2}\right)^{\frac{3}{2}}} \right],
\end{aligned}
\end{equation}
where we approximated $\delta\mu^{2} \approx 0$ in the last line, and introduced the lower cut-off $\alpha$. Factorizing out the factors of $k^{2}$, the fraction can be simplified to 
\begin{equation}
    \frac{k^{2}}{\left( \frac{k^{4}}{4} - k^{2}\delta\mu + 4k^{2}\right)^{\frac{3}{2}}} = \frac{1}{|k|\left( \frac{k^{2}}{4} - \delta\mu + 4\right)^{\frac{3}{2}}} \approx \frac{1}{|k|\left( \frac{k^{2}}{4} + 4\right)^{\frac{3}{2}}} +\mathcal{O}\left(\delta\mu\right)
\end{equation}
where in the last step we made a Taylor expansion for $\delta\mu$ close to zero. Thus, at the QCP the first integral diverges with $|k|^{-1}$ in the integrand, and a quick computation with WolframAlpha shows 
a logarithmic divergence proportional to 
\begin{equation}
 \sim -\frac{L}{2\pi} \log\left( \frac{\epsilon^{2}}{\alpha^{2}}\right) + \mathcal{O}\left(\epsilon^{2}\right).
\end{equation}\\
{\color{blue}\subsection{Probe of BCS Pairing Effective Strength}
}

In the low-frequency limit, the transition rates in Eq. (14) of the Letter also probe the effective BCS pairing strength defined by 
\begin{equation}
\Delta_{BCS} = \int_{0}^{\pi}\text{d}k \sin\left(\theta_{k}\right). 
\end{equation}
In the $\omega \longrightarrow 0$ limit and to linear order in $\epsilon$ we find the expression (for $\phi = 0$)
\begin{equation}
  \frac{\mathcal{P}_{+} + \mathcal{P}_{-}}{2} =  \frac{B^{2}_{0}L^{2}}{\Delta^{2}\pi^{2}}\left|\int_{0}^{\pi}\text{d}k \sin\left(\theta_{k}\right)\right|^{2} + \mathcal{O}\left(\omega^{2}\right).
\end{equation}
We used $\sin\left(k\right)/E\left(k\right) \sim \sin\left(\theta_{k}\right)$ according to the mapping on the sphere. The transition probabilities therefore also present a protocol to measure the effective pairing strength $\Delta_{BCS}$, which also maps the topological phase transition as proposed in Ref. [13] of the Letter. The limit $\omega \approx 0$ could be extracted perturbatively or by a ``quench" at time $t = 0$ of the SC pairing amplitude $\Delta$.\\

{\color{blue}\subsection{Capacitance Probe for the Topological Phase Transition}}

For $\phi = 0$, the current operator $\mathcal{J}_{\Delta}$ is given in terms of the quasi-particle operators $O^{\pm}$ and $O^{z}$ as 
\begin{equation}
\mathcal{J}_{\Delta} = {\color{black}4}\Delta \sum_{q} \sin\left(k\right)\left(O_q^{+} + O_q^{-}\right)
\end{equation}
whilst
the charge reads
\begin{equation}
\mathcal{Q} = - \sum_{k}\left( \cos\left(\theta_{k}\right)O^{z}_{k} + i\sin\left(\theta_{k}\right)\left(O^{+}_{k} - O^{-}_{k}\right)\right).
\end{equation} 
Due to $O_{q}^{-}|BCS\rangle = 0$, only the $\left\langle O_{q}^{-}O_{k}^{+} \right\rangle_{0}\sim \delta\left(k- q\right)$ contributes non-trivially to the response $\Delta \mathcal{Q}_{L}\left(t\right)$. Together with $\omega^{\pm}_{k} = \omega \pm 2E\left(k\right)$, we find
\onecolumngrid
\begin{equation}\label{charge_time}
\begin{aligned}
       \Delta\mathcal{Q}_{L}\left(t\right)=&\ {\color{black} -} i\frac{B_{0}L^{2}}{4\pi^{2}}\int_{0}^{\pi}\text{d}k\int_{0}^{\pi}\text{d}q \int_{0}^{t}\text{d}\tau\ \sin\left(q\right)\sin\left(\theta_{k}\right)\cos\left(\omega \tau\right) \left\langle  i O^{-}_{q} e^{iH_{0}\left(t - \tau\right)} O^{+}_{k} - i O_{k}^{-} e^{-iH_{0}\left(t - \tau\right)} O^{+}_{q}\right\rangle_{0} \\ 
      =& {\color{black} +}\frac{B_{0}L}{2\pi}\int_{0}^{\pi}\text{d}k \sin\left(k\right)\sin\left(\theta_{k}\right) \left[ \frac{\sin\left(\omega t\right)  - \sin\left(2E(k)t\right)}{\omega^{+}_{k}} + \frac{\sin\left(\omega t\right)  + \sin\left(2E(k)t\right)}{\omega^{-}_{k}}\right].
\end{aligned}
\end{equation}
There are two competing oscillations in \eqref{charge_time} above, and we perform an additional time-average over the domain $t \in \left[0, \frac{2\pi}{\omega}\right]$ to remove the oscillations in $\omega t$. Thus, the low-frequency limit will correspond to an infinite-time averaging of the linear charge fluctuations induced by $\mathcal{A}$. The low-frequency limit can then be achieved by either a quench of $\Delta\left(t\right) = \Delta + \theta\left(t\right)\delta \Delta$, or through ultra-low frequency electromagnetic waves. Alternatively, a perturbative expansion in $\omega$ and subsequent scaling analysis could also yield similar results. We find 
\begin{equation}\label{tildeQ}
     \tilde{\mathcal{Q}}_{L}\left(\omega\right) = {\color{black}-}\frac{1}{T}\int_{0}^{T = \frac{2\pi}{\omega}}\Delta \mathcal{Q}_{L}\left(t\right)
       = - \frac{B_{0}L\omega}{4\pi^{2}}\int_{0}^{\pi}\text{d}k \frac{\sin\left(k\right)\sin\left(\theta_{k}\right)}{2E\left(k\right)}\sin^{2}\left(\frac{2\pi E\left(k\right)}{\omega}\right)\frac{4E\left(k\right)}{\omega^{2} - 4E^{2}\left(k\right)}.
\end{equation}

\end{document}